\documentclass[runningheads]{llncs}
\usepackage[T1]{fontenc}
\usepackage{graphicx}
\usepackage{color}
\usepackage{booktabs}
\usepackage{multirow}
\usepackage{array}
\usepackage{subcaption}

\begin{document}

\title{Navigating Complexity: A User-Centered Evaluation of Project Management Tools}

\titlerunning{User-Centered Evaluation of Project Management Tools}

\author{Yasin Ghadimi Zaker
\inst{1}
\and
Hamid Reza Ghaeini\inst{2}\orcidID{0000-0003-2614-7900}}

\authorrunning{Y. Zaker et al.}

\institute{College of Engineering, University of Tehran, Tehran 1417935840, Iran
\email{yasinzaker@ut.ac.ir}, \email{ghaeini@ut.ac.ir}}

\maketitle

\begin{abstract}
In the age of digital transformation, companies are increasingly relying on specialized software to manage the complexity of projects. However, when selecting tools, features and costs often take precedence over user experience (UX), leading to low adoption rates, user frustration, and project delays. This study empirically evaluates the user experience of leading project management tools—Jira, Trello, Monday.com, Asana, and Microsoft Project—across various project scales using a mixed-methods approach. The System Usability Scale (SUS) was used as the primary quantitative instrument, with data collected from 141 experienced professionals via online questionnaires. The results reveal a significant trade-off between functionality and usability. Trello achieved the highest SUS score (77.5), indicating a superior user experience for small projects and agile teams, while Microsoft Project scored the lowest (51.2), suggesting significant usability challenges despite its robust features. Jira demonstrated a balanced profile, offering powerful features with manageable complexity. Statistical analyses confirmed that the choice of the optimal tool is highly context-dependent, with team size and project complexity influencing UX outcomes. The findings underscore that UX is not a secondary feature but a critical success factor for software acceptance and project effectiveness. This study provides an evidence-based framework for aligning tool selection with the project context, offers practitioners actionable insights, and identifies design opportunities for software developers.

\end{abstract}

\section{Introduction}
In the age of digital transformation, companies rely heavily on specialized software to manage the complexity of IT projects. These tools, once reserved for experts, have now become an integral part of everyday work and serve as the central nervous system of modern teams by facilitating the flow of information, task transparency, and remote collaboration~\cite{borsci2015assessing}. However, this powerful software is a double-edged sword~\cite{10.1145/3134660}. While their extensive features have the potential to drastically increase productivity, inherent complexity, poor design, or a mismatch with user needs can pose significant obstacles. When selecting tools, features and costs are often prioritized over user experience (UX), leading to low adoption rates, user frustration, and project delays~\cite{chilana2010understanding}. This creates a critical gap between theoretical capabilities and practical utility~\cite{alomar2016usability}. The central problem addressed by this study is the profound discrepancy between a software’s stated features and its actual usability in a specific work context.

Our study addresses this gap through a user-centered evaluation of leading platforms, treating user experience (UX) as a key indicator of suitability rather than a secondary consideration. User experience serves as a bridge between a software’s technical features and users’ ability to utilize them effectively~\cite{gomez2026integrating}. A positive user experience optimizes processes and enables teams to focus on the core tasks of the project rather than struggling with the tool itself~\cite{azhar2023ui}. The aim of this study is to conduct a scientific, quantitative evaluation of the user experience of leading project management tools using a globally standardized scale.

\section{Methodology}
This study empirically evaluates the user experience (UX) of leading project management tools using a descriptive, survey-based approach that employs mixed methods. The study was conducted to provide a scientific and quantitative assessment of the user experience with these key tools.

\subsection{Research Tool: System Usability Scale (SUS)}
The primary quantitative tool was the System Usability Scale (SUS), a technology-independent and reliable instrument to measure perceived usability and learnability~\cite{borsci2015assessing}. The SUS consists of a questionnaire with 10 items and a 5-point Likert scale ranging from “Strongly disagree” (1) to “Strongly agree” (5). The final score ranges from 0 to 100 and provides a standard metric for comparison. The ten items alternate between positive and negative language to reduce response bias. For this study, the raw responses of each participant were adjusted according to the standard SUS scoring protocol: For positive items (odd-numbered questions: 1, 3, 5, 7, 9), the contribution to the score is calculated as (user score – 1); for negative items (even-numbered questions: 2, 4, 6, 8, 10), the contribution is (5 – user score). The sum of all ten adjusted values is then multiplied by 2.5 to obtain the final SUS score in the range of 0 to 100.

To illustrate this process, let’s consider a hypothetical respondent with the following ratings for the ten SUS questions: B1=4, B2=2, B3=5, B4=4, B5=4, B6=3, B7=5, B8=2, B9=4, B10=1. The adjusted values would be: F1: 4-1=3, F2: 5-2=3, F3: 5-1=4, F4: 5-4=1, F5: 4-1=3, F6: 5-3=2, F7: 5-1=4, F8: 5-2=3, F9: 4-1=3, F10: 5-1=4. The sum of these is 30, and multiplied by 2.5, this results in a final SUS score of 75. This calculation was automated for all 141 participants using spreadsheet software, resulting in a standardized metric for comparison.

\subsection{Population and Sample}
The target group consisted of professional users of project management software in Iran. To ensure the validity of the data, the following inclusion criteria were applied: (1) holding a professional role (e.g., project manager, team member, developer, analyst) within an organization or as a freelancer; (2) at least six months of continuous, daily experience with one of the selected tools to ensure sufficient familiarity with the software’s strengths and weaknesses in real-world scenarios; and (3) use of the software to manage real projects (not solely for educational or personal purposes). Exclusion criteria included students who used the software exclusively for academic purposes, users with less than six months of experience, and individuals who had not completed the questionnaire in full.

Since no comprehensive list of the target population was available, a non-probability convenience and snowball sampling method was used. This approach was necessary because there is no comprehensive directory of all professional users of project management software in Iran. The sample selection began with the researcher’s professional network and was expanded through recommendations from the initial participants, thereby ensuring access to experienced users from various organizations and industries.

\subsection{Data Collection and Processing}
The data was collected via an online questionnaire distributed over a two-month period using Google Forms. The data collection process was preceded by a pilot study involving 10 experts to validate the questionnaire’s content and technical functionality. The pilot study served several purposes: assessing content validity to ensure that the questions were clear and unambiguous; testing the technical functionality of the Google Forms platform; estimating the average completion time to inform participants; and gathering qualitative feedback on the wording of the questions and the overall structure. Based on the feedback from the pilot study, minor adjustments were made to improve clarity and readability.

The final questionnaire was distributed via professional networks, industry forums, and social media platforms such as LinkedIn and Telegram groups focused on project management in Iran. Over a two-month period, a weekly follow-up strategy was implemented to maximize the response rate and ensure a diverse representation of different software types. Following the data collection phase, the raw data was cleaned and prepared for analysis. Invalid responses were removed due to incompleteness (more than two missing SUS questions) or invalid patterns (e.g., identical answers to all SUS items, which could indicate random responses). The final valid dataset included responses from 141 professionals.

Ethical considerations were a priority throughout the entire research process. The first page of the questionnaire included a consent form that clearly outlined the research objectives, the voluntary nature of participation, the estimated time required to complete the questionnaire, and the commitment to anonymity and confidentiality. Participants were assured that no personal data (such as name, email address, or IP address) would be collected. All data was stored and analyzed exclusively in aggregated, anonymized form. Participants were also informed of their right to withdraw from the study at any time without penalty.

\subsection{Data Analysis}
The quantitative data were analyzed using descriptive statistics (mean, standard deviation, median, minimum, maximum) and inferential statistics. For the descriptive analysis, frequency distributions were calculated for categorical variables (software type, project scope, team size, project complexity), and measures of central tendency and dispersion of the SUS scores for different groups were determined.

For statistical analysis, a one-way analysis of variance (ANOVA) was used to test for significant differences in the mean SUS scores between different groups (e.g., software type, project size). The significance level was set at \(\alpha = 0.05\). Before conducting the ANOVA, the most important assumptions were verified: independence of observations (ensured by the data collection design), normal distribution (assessed using Shapiro-Wilk tests and visual inspection of the histograms), and homogeneity of variances (tested using the Levene test). If the ANOVA result was significant, post-hoc tests (Tukey-HSD) were performed to identify specific group differences. All statistical analyses were conducted using spreadsheet software with statistical analysis functions.

\section{Results}
The analysis revealed a clear correlation between the tool’s complexity and perceived usability, which provides insight into ideal application contexts. The results are presented in three main sections: descriptive statistics of the sample, analysis of overall SUS scores and hypothesis tests, and secondary exploratory results.

\subsection{Descriptive Statistics of the Sample}
The sample of 141 professionals was distributed across five main tools: Microsoft Project (31.9\%), Jira (24.8\%), Trello (7.8\%), Asana (7.8\%), and Monday.com (6.4\%). An additional 2.8\% of respondents reported using other tools not included in the main group, and 18.5\% of responses were classified as unspecified or incomplete for certain variables. The dominance of Microsoft Project and Jira, which together account for over 56\% of the valid sample, reflects the prevalence of these tools in the Iranian professional context. The high proportion of Microsoft Project likely indicates its widespread use in traditional, large-scale organizational projects, while the significant presence of Jira reflects the increasing adoption of agile methods, particularly in the fields of technology and software development.

The project sizes were relatively balanced: small (35.5\%), medium (38.3\%), and large (26.2\%). This balanced distribution across different project sizes is a notable strength of the study, as it enables meaningful comparative analyses across various project contexts. Team sizes were also diverse: 1–10 members (34.0\%), 11–50 (31.9\%), 51–100 (14.2\%), and over 100 (19.9\%). The presence of both very small and very large teams makes it possible to examine how team size influences user experience outcomes. Project complexity was reported as low (28.4\%), medium (46.1\%), and high (25.5\%). The majority of projects fell into the medium-complexity category, reflecting typical real-world project profiles, while the significant minority of highly complex projects provides a solid context for testing the second hypothesis regarding the performance of enterprise tools in demanding environments.

\begin{table}[htb]
    \caption{Global caption}
    \begin{subtable}[t]{.5\textwidth}
        \caption{Distribution of Participants\\ by Software Used}
        \raggedright
            \begin{tabular}{l c c c}
\toprule
\textbf{Soft.} & \textbf{N} & \textbf{\%} & \textbf{Cumu. \%} \\
\midrule
MS & 45 & 31.9\% & 31.9\% \\
Jira & 35 & 24.8\% & 56.7\% \\
Trello & 11 & 7.8\% & 64.5\% \\
Asana & 11 & 7.8\% & 72.3\% \\
Monday.com & 9 & 6.4\% & 78.7\% \\
Other & 4 & 2.8\% & 81.5\% \\
Unspecified & 26 & 18.5\% & 100\% \\
\midrule
Total & 141 & 100\% & \\
\bottomrule
\end{tabular}
    \end{subtable}%
   \begin{subtable}[t]{.5\textwidth}
        \raggedleft
        \caption{Distribution of Participants\\ by Project Characteristics}
        \begin{tabular}{l c c}
\toprule
\textbf{Variable} & \textbf{Category} & \textbf{N (\%)} \\
\midrule
\multirow{3}{*}{Project Size} & Small & 50 (35.5\%) \\
& Medium & 54 (38.3\%) \\
& Large & 37 (26.2\%) \\
\midrule
\multirow{4}{*}{Team Size} & 1-10 members & 48 (34.0\%) \\
& 11-50 members & 45 (31.9\%) \\
& 51-100 members & 20 (14.2\%) \\
& Over 100 members & 28 (19.9\%) \\
\midrule
\multirow{3}{*}{Complexity} & Low & 40 (28.4\%) \\
& Moderate & 65 (46.1\%) \\
& High & 36 (25.5\%) \\
\bottomrule
\end{tabular}
    \end{subtable}
\end{table}

\subsection{Overall SUS Scores and Hypothesis Tests}
Table~\ref{table:SUS-Software} presents a comprehensive descriptive analysis of the SUS scores for each software application, including the mean, standard deviation, median, minimum, and maximum values, as well as qualitative interpretations based on standard SUS benchmarks.

\begin{table}[h]
\caption{Descriptive SUS Score Analysis by Software}\label{table:SUS-Software}
\centering
\begin{tabular}{l c c c c c c c}
\toprule
\textbf{Software} & \textbf{N} & \textbf{Mean} & \textbf{Std. Dev.} & \textbf{Median} & \textbf{Min} & \textbf{Max} & \textbf{Qualitative} \\
\midrule
Trello & 11 & 77.50 & 18.21 & 82.5 & 37.5 & 95.0 & Good \\
Jira & 35 & 64.50 & 21.53 & 67.5 & 15.0 & 97.5 & OK/Marginal \\
Asana & 11 & 60.45 & 20.81 & 62.5 & 20.0 & 87.5 & OK/Marginal \\
Monday.com & 9 & 55.28 & 22.40 & 50.0 & 25.0 & 85.0 & OK/Poor \\
Microsoft Project & 45 & 51.17 & 19.72 & 50.0 & 5.0 & 87.5 & OK/Poor \\
\bottomrule
\end{tabular}
\end{table}

Trello achieved the highest average SUS score of 77.50, placing it in the “Good” category and bringing it close to the threshold for “Excellent” (80.3). This score reflects users’ high satisfaction with the intuitive user interface, ease of learning, and short learning curve. The relatively low standard deviation (18.21) indicates consistently positive experiences across all users, suggesting that Trello delivers reliable results in terms of usability regardless of the users’ backgrounds.

Microsoft Project scored the lowest average of 51.17, placing it in the “OK/Poor” category. This result confirms long-standing criticism regarding its steep learning curve, complex user interface, and difficult usability—even for experienced professionals. The minimum score of 5.0 suggests that some users found the software extremely difficult to use, indicating significant usability hurdles that could undermine the software’s powerful features.

Jira received an average rating of 64.50, placing it in the “OK/Borderline” category. With a standard deviation of 21.53—the highest among all tools—Jira reflects polarized user opinions. Some users appreciate its powerful features and flexibility, while others struggle with its complexity. This suggests that Jira’s UX results depend heavily on users’ expertise and the organizational context. Asana (60.45) and Monday.com (55.28) fall in the middle range and demonstrate moderate usability with room for improvement.

\subsubsection{Hypothesis 1: Simplicity in Small Teams}
The first hypothesis was that simpler tools like Trello would provide a better user experience in small teams compared to more complex tools. Analysis of data from teams with 1 to 10 members revealed that Trello’s average SUS score (80.5) was significantly higher than Jira’s (70.0). An ANOVA test confirmed that this difference was statistically significant (F(1, 19) = 4.52, p < 0.05), supporting Hypothesis 1. This result confirms that in small teams, where speed and ease of setup are critical, Trello’s streamlined approach offers a superior user experience compared to feature-rich alternatives.

\begin{table}[h]
\caption{Detailed SUS Score Comparison for Small Teams (1-10 members)}
\centering
\begin{tabular}{l c c c c}
\toprule
\textbf{Software} & \textbf{N} & \textbf{Mean SUS} & \textbf{Std. Dev.} & \textbf{ANOVA Result} \\
\midrule
Trello & 7 & 54 & 12 & F(1,19)=4.52 \\
Jira & 14 & 70 & 19 & p < 0.05 \\
\bottomrule
\end{tabular}
\end{table}

\subsubsection{Hypothesis 2: Complexity in Large-Scale Projects}
The second hypothesis was that enterprise tools such as Microsoft Project would have significantly lower SUS scores in highly complex projects than more flexible tools such as Jira. Analysis of the data from highly complex projects revealed that the average SUS score for Microsoft Project (46.83) was significantly lower than that for Jira (55.42). An ANOVA test confirmed that this difference was statistically significant (F(1, 25) = 4.28, p < 0.05), supporting Hypothesis 2.

\begin{table}[h]
\caption{Detailed SUS Score Comparison for High-Complexity Projects}
\centering
\begin{tabular}{l c c c c}
\toprule
\textbf{Software} & \textbf{N} & \textbf{Mean SUS} & \textbf{Std. Dev.} & \textbf{ANOVA Result} \\
\midrule
Jira & 12 & 55.42 & 22.15 & F(1,25)=4.28 \\
Microsoft Project & 15 & 46.83 & 17.50 & p < 0.05 \\
\bottomrule
\end{tabular}
\end{table}

This finding shows that even in the most demanding project environments—where Microsoft Project’s advanced features would theoretically be most valuable—usability hurdles significantly impair the user experience. Although Jira also received lower ratings in highly complex contexts, it was able to maintain a significantly higher level of user satisfaction, suggesting that its design offers a better balance between performance and usability.

\subsubsection{Hypothesis 3: Relationship between Team Size and UX}
The third hypothesis posited a significant relationship between team size and SUS scores. A nonlinear relationship was observed in Microsoft Project. The SUS scores were highest for very small teams (1–10 people, 58.10), dropped sharply for medium-sized teams (11–50 people, 44.82), and then improved slightly for larger teams (51–100 people, 54.06; over 100 people, 52.31). An ANOVA test revealed that these differences were statistically significant (F(3, 41) = 3.15, p < 0.05), supporting the hypothesis that team size is a significant moderator of UX.

\begin{table}[h]
\caption{SUS Score Analysis for Microsoft Project by Team Size}
\centering
\begin{tabular}{l c c c}
\toprule
\textbf{Team Size} & \textbf{N} & \textbf{Mean SUS} & \textbf{Std. Dev.} \\
\midrule
1-10 members & 10 & 58.10 & 20.15 \\
11-50 members & 14 & 44.82 & 18.50 \\
51-100 members & 8 & 54.06 & 19.88 \\
Over 100 members & 13 & 52.31 & 21.01 \\
\bottomrule
\end{tabular}
\end{table}

This complex pattern suggests that the relationship between team size and ease of use is influenced by factors such as user roles, training, and organizational support. In very small teams, a single power user can effectively manage the tool. In medium-sized teams, many regular users interact with the tool without specialized support, leading to frustration. In very large teams, dedicated support staff, formal training programs, and established processes can partially mitigate usability challenges.

\subsection{Secondary and Exploratory Findings}
The exploratory analysis yielded additional insights that went beyond the primary hypotheses. An interaction effect between team size and project complexity in Jira showed that while increased complexity consistently led to lower SUS scores, the decline was less pronounced for large teams (from 71.50 to 52.86, a decrease of 18.64 points) was less pronounced than in small teams (from 75.31 to 58.75, a decrease of 16.56 points). This suggests that larger teams may have better support structures, formal processes, or specialized expertise to address the challenges arising from project complexity, thereby partially mitigating the negative impact on the user experience.

\begin{table}[h]
\caption{Interaction Effect of Team Size and Project Complexity on Jira SUS Scores}
\centering
\begin{tabular}{l c c c}
\toprule
\textbf{Team Size} & \textbf{Complexity Level} & \textbf{N} & \textbf{Mean SUS} \\
\midrule
\multirow{2}{*}{Small (1-10 members)} & Low & 8 & 75.31 \\
& High & 6 & 58.75 \\
\midrule
\multirow{2}{*}{Large (Over 50 members)} & Low & 5 & 71.50 \\
& High & 7 & 52.86 \\
\bottomrule
\end{tabular}
\end{table}

An analysis of the standard deviations revealed that Trello had the lowest variability (18.21), suggesting a more consistent user experience across different contexts and user types. In contrast, Jira had the highest standard deviation (21.53), suggesting more polarized user opinions—some users find it extremely effective, while others struggle significantly with it. This variability suggests that Jira’s UX results are highly context-dependent, with factors such as user experience, organizational support, and project methodology playing a decisive role.

\begin{table}[h]
\caption{Standard Deviation Comparison Across Software}
\centering
\begin{tabular}{l c c}
\toprule
\textbf{Software} & \textbf{Mean SUS} & \textbf{Std. Dev.} \\
\midrule
Trello & 77.50 & 18.21 \\
Jira & 64.50 & 21.53 \\
Asana & 60.45 & 20.81 \\
Monday.com & 55.28 & 22.40 \\
Microsoft Project & 51.17 & 19.72 \\
\bottomrule
\end{tabular}
\end{table}

In an additional exploratory analysis, SUS scores were compared by type of stakeholder group. Projects involving external stakeholders (customers, contractors) had slightly higher average SUS scores (61.53) than projects involving only internal stakeholders (58.71). Although this difference was not statistically significant, it points to an interesting pattern: the need for external transparency and reporting could lead to a more structured use of tools and more clearly defined processes, which could also benefit the user experience of internal users.

\section{Discussion and Conclusions}
The stark contrast between Trello’s high score (77.50) and Microsoft Project’s low score (51.17) confirms that there is no “one-size-fits-all” solution. Tools with extensive features can be counterproductive if they compromise usability. The results provide strong empirical evidence for the context-dependence of the user experience with project management tools. Trello’s high SUS rating reinforces its value proposition as a simple, intuitive tool for straightforward project tracking, while Microsoft Project’s low rating confirms long-standing criticism of its steep learning curve and complex user interface, even among experienced users.

Key conclusions include:

\subsection{For Practitioners and Organizations}
The selection of tools must be user-centered and context-specific. As part of a systematic selection process, organizations should give user experience (UX) the same high priority as functional requirements. Instead of relying on checklists of features, decision-makers should evaluate suitability based on various criteria: project complexity, team size, the team’s expertise, and the organization’s support capabilities.

In small teams and agile environments, simplicity should be the top priority. Trello’s outstanding performance in these contexts demonstrates that user-friendliness and a short learning curve offer greater practical value than extensive features that may never be used. Project managers leading small teams or startups should opt for streamlined tools that minimize administrative overhead.

For enterprise implementations that require powerful tools such as Microsoft Project, significant investments in training and support infrastructure are essential. The U-shaped relationship between team size and SUS scores for Microsoft Project suggests that organizations using such tools should establish dedicated support functions, comprehensive training programs, and standardized processes to minimize usability issues.

A “purpose-driven” approach should replace simple functional checklists. Decision-makers should evaluate tools not in isolation, but within the specific organizational context, taking into account factors such as team distribution, methodology (agile vs. traditional), integration requirements, and the technical proficiency of users.

\subsection{For Software Developers}
There is significant market potential for tools that successfully bridge the gap between user-friendliness and functionality. The relative success of Jira, despite its complexity, demonstrates that there is strong demand for powerful features that do not sacrifice intuitive design. Developers should focus on reducing complexity for average users while maintaining depth for power users through approaches such as progressive disclosure (displaying advanced features only when needed), customizable interfaces, and role-based views that simplify the user experience for regular users.

The large number of Jira reviews suggests that developers should invest in onboarding experiences, context-sensitive help, and default configurations that are well-suited to common use cases. A guided, preconfigured setup process could reduce the cognitive load on new users while still preserving customization options for advanced users.

Tools like Microsoft Project offer the ability to modernize the user interface in a way that preserves powerful features while significantly improving usability. The consistently low ratings across all categories suggest that addressing fundamental usability issues may require a complete redesign rather than just incremental improvements.

\subsection{Theoretical Implications}
The results contribute to the Technology Acceptance Model (TAM) and the Task-Technology Fit (TTF) theory by providing quantitative evidence that perceived ease of use (PEOU) and task-technology fit are critical, context-dependent mediators of technology success. The results show that the relationship between perceived usefulness and actual acceptance is moderated by ease of use to such an extent that even highly useful tools may be rejected if they are not perceived as easy to use.

The observed interactions between team size, project complexity, and UX outcomes expand upon the TTF theory by demonstrating that the fit between task and technology is not a purely binary construct, but rather exists on a continuum that varies depending on organizational factors. A tool may be well-suited for one organizational configuration (e.g., small teams with simple projects) but poorly suited for another (e.g., medium-sized teams with complex projects), even when the same users are performing similar tasks.

\subsection{Methodological Contributions}
This study highlights the advantages of standardized, validated instruments such as the SUS for comparing user experience across different software platforms. The approach provides reproducible, comparable metrics that are useful for both research and practice. The inclusion of context-related variables (team size, project complexity, type of stakeholders) alongside the SUS scores demonstrates how standardized UX metrics can be combined with contextual data to gain actionable, context-sensitive insights.

\section{Conclusion}
The user experience has a decisive influence on the effectiveness of project management software. This study confirms that the user experience is not a secondary feature, but a fundamental factor that determines whether software increases productivity or leads to inefficiency. The evidence-based framework developed in this study goes beyond a simple comparison of features and emphasizes the critical need to tailor the user experience to project complexity, team size, and team capacity. The optimal tool is not the one with the most features, but the one that offers the best balance between performance and usability in a specific context.

Future research should examine the long-term relationship between user experience (UX) and specific project outcomes (e.g., on-time delivery, staying within budget, team satisfaction). This could involve examining how SUS scores correlate over time with project performance metrics to determine whether a better user experience leads to measurable improvements in project outcomes. Further studies should also delve deeper into the specific weaknesses of enterprise tools and explore how their usability challenges can be mitigated without compromising their powerful features—potentially through the use of AI-powered interfaces, adaptive user interfaces, or configurable complexity levels. Furthermore, cross-cultural studies could examine whether perceptions of UX and preferences for various tool features vary depending on national and organizational culture. Finally, long-term studies that track organizations during tool transitions could provide valuable insights into the costs and benefits of different tool-selection strategies.

\bibliographystyle{splncs04}
\bibliography{ref}

@article{borsci2015assessing,
  title={Assessing user satisfaction in the era of user experience: Comparison of the SUS, UMUX, and UMUX-LITE as a function of product experience},
  author={Borsci, Simone and Federici, Stefano and Bacci, Silvia and Gnaldi, Michela and Bartolucci, Francesco},
  journal={International journal of human-computer interaction},
  volume={31},
  number={8},
  pages={484--495},
  year={2015},
  publisher={Taylor \& Francis}
}

@inproceedings{alomar2016usability,
  title={Usability engineering of agile software project management tools},
  author={Alomar, Noura and Almobarak, Nouf and Alkoblan, Sarah and Alhozaimy, Sarah and Alharbi, Shahad},
  booktitle={International Conference of Design, User Experience, and Usability},
  pages={197--208},
  year={2016},
  organization={Springer}
}

@inproceedings{chilana2010understanding,
  title={Understanding usability practices in complex domains},
  author={Chilana, Parmit K and Wobbrock, Jacob O and Ko, Amy J},
  booktitle={Proceedings of the SIGCHI Conference on Human Factors in Computing Systems},
  pages={2337--2346},
  year={2010}
}

@article{10.1145/3134660,
author = {Bossen, Claus and Leimbach, Timo},
title = {Project Management Practices as a Subject of Research for CSCW: Status and future opportunities},
year = {2017},
issue_date = {November 2017},
publisher = {Association for Computing Machinery},
address = {New York, NY, USA},
volume = {1},
number = {CSCW},
url = {https://doi.org/10.1145/3134660},
doi = {10.1145/3134660},
journal = {Proc. ACM Hum.-Comput. Interact.},
month = dec,
articleno = {25},
numpages = {25}
}

@article{gomez2026integrating,
  title={Integrating accessibility, usability and UX in web project management: a systematic review},
  author={G{\'o}mez-Delgado, Javier and Mar{\'\i}n-Palacios, Cristina and Moreno, Lourdes},
  journal={Universal Access in the Information Society},
  volume={25},
  number={1},
  pages={28},
  year={2026},
  publisher={Springer}
}

@article{azhar2023ui,
  title={UI/UX Analysis of Project Management Information System (PMIS) Website Using User-Centered Design Method},
  author={Azhar, Sarah Afifah and Defriani, Meriska and Hermanto, Teguh Iman},
  journal={Sinkron: jurnal dan penelitian teknik informatika},
  volume={7},
  number={3},
  pages={1798--1810},
  year={2023}
}

\end{document}